\documentstyle[12pt]{article}

\newcommand{\newsection}{
\setcounter{equation}{0}
\section}
\newcommand{\tr}[1]{\,{\rm tr}\,#1\,}
\def\e{{\,\rm e}\,}

\def\eop{\vspace*{\fill}\pagebreak}
\def\be{\begin{equation}}
\def\ee{\end{equation}}
\def\bea{\begin{eqnarray}}
\def\eea{\end{eqnarray}}

\def\Mdclose#1{{\bar {\cal M}}^{disc}_{#1}}

\def\L{\Lambda}
\def\SS{\Sigma}
\def\l{\lambda}

\def\ep {\epsilon}

\def\Mcomb{{\cal M}_{g,n}^{comb}}
\def\Mdisc{{\cal M}_{g,n}^{disc}}
\def\Mcdisc{{\bar {\cal M}}_{g,n}^{disc}}
\def\MgnR{{\cal M}_{g,n}\times {\bf R}_{+}^{n}}
\def\Mgn{{\cal M}_{g,n}}
\def\Mc{{\bar {\cal M}}_{g,n}}

\newcommand{\dd}[1]{{\partial \over \partial #1}}

\newcommand{\p}{^{\prime}}
\newcommand{\ra}{\rightarrow}

\newcommand{\OO}{{\cal O}}
\newcommand{\fr}[2]{{\textstyle {#1 \over #2}}}

\title{{\bf \mbox{} \\Matrix Model for Discretized Moduli Space}
\vspace{.5cm}}
\author{{\bf L. Chekhov}\thanks{E--mail: \ chekhov@qft.mian.su}
\date{ }
\vspace{.5cm} \\
{\it Steklov Mathematical Institute} \\
{\it Vavilov st.42, GSP-1, 117966 Moscow, Russia}}

\begin{document}

\maketitle

\vspace{-9.6cm}

\begin{flushright}
SMI--92--04 \\ May 11, 1992
\end{flushright}

\vspace{6.8cm}

\begin{abstract}
We study the algebraic geometrical background of the Penner--Kontsevich
matrix model with the potential
$N\alpha \tr {\bigl(- \fr 12 \L X\L X +\log (1-X)+X\bigr)}$. We show that this
model describes intersection indices of linear bundles on the discretized
moduli space right in the same fashion as the Kontsevich model is related
to intersection indices (cohomological classes) on the Riemann surfaces of
arbitrary genera. The special role of the logarithmic potential originated
from the Penner matrix model is
demonstrated. The boundary effects which was unessential in the case of the
Kontsevich model are now relevant, and intersection indices on the
discretized moduli space of genus $g$ are expressed through Kontsevich's
indices of the genus $g$ and of the lower genera.
\end{abstract}

\eop

\newsection{Introduction.}

The last three years of development in matrix models initiated by the papers
\cite{BK90} revealed a lot of applications of these models in various branchs
of mathematical physics: two--dimensional quantum field theory, intersection
theory on the moduli space of Riemann surfaces, integrable hierarchies,
matrix integrals, random surfaces and others. The approach by \cite{BK90}
where the explicit solution was presented deal with triangulated Riemann
surfaces where any triangulation determines some singular metric obtained by
the arrangement of equilateral triangles. One can think that when the number
of triangles tends to infinity these singular metrics approximate ``random
metrics'' on the surface. These triangulations were presented by a
hermitean $N\times N$ one--matrix model
\be
\int \exp\bigl( \tr P(X) \bigr) DX,
\label{herm}
\ee
where $P(X)=\sum _{n}T_n \tr X^n$, $T_n$ being times for the
one--matrix model.
For this system discrete Toda chain equations holds with an additional
Virasoro symmetry imposed \cite{KawMMM}. In the limit $N\to\infty$ the
Korteveg--de--Vries equation arises. The partition function of the
two--dimensional gravity for this approach is a series in an infinite number
of variables and coincides with the logarithm of some $\tau$--function for
KdV hierarchy.

Another approach to the two--dimensional gravity is to do the integral over
all classes of conformally equivalent metrics on Riemann surfaces. It may be
presented as an integral over the finite--dimensional space of conformal
structures. This integral has a cohomological description as an intersection
theory on the compactified moduli space of complex curves. Edward Witten
presented compelling evidence for a relationship between random surfaces and
the algebraic topology of moduli space \cite{Wit1}, \cite{Wit2}. In fact, he
suggested that these expressions coincide since both satisfy the same
equations of KdV hierarchy. It was Maxim Kontsevich who proved this
assumption \cite{Kon91}. Surprisingly, he explicitly presented a new matrix
model defining exactly the values of intersection indices or, on the language
of 2D gravity, correlation functions of observables $\OO _n$ of the type
\be
<\OO _{n_1}\dots \OO_{n_s}>_g,
\ee
where $<\dots >_g$ denotes the expectation value on a Riemann surface with
$g$ handles. Then the string partition function $\tau (t)$ has an asymptotic
expansion of the form
\be
\tau (t)=\exp \sum_{g=0}^{\infty}\left\langle \exp \sum_{n}t_n \OO _n
\right\rangle_{g},
\label{taufunct}
\ee
and it is a tau--function of the KdV hierarchy taken at a point of
Grassmannian where it is invariant under the action of the set of the
Virasoro constraints: ${\cal L}_n \tau (t)=0$, $n\ge -1$ \cite{Fuk},
\cite{Wit3}, \cite{GN}, \cite{MMM}. One might say that the Kontsevich model
is used to triangulate moduli space, whereas the original models triangulated
Riemann surfaces (see e.g. \cite{Dij91}).

The generalization of the Kontsevich model --- so-called Generalized
Kontsevich Model (GKM) \cite{KMMMZ} is related to the two--dimensional Toda
lattice hierarchy and it originated from the external field problem defined
by the integral
\be
Z[\L ;N]=\int DX \exp \left\{N \tr {\bigl( \L X -V_0 (X)\bigr)}\right\},
\label{external}
\ee
where $V_0(X)=\sum_{n} t_n \tr X^n$ is some potential, $t_n$ are related to
times of the hierarchy. This model is equivalent to the Kontsevich integral
for $V_0(X) \sim \tr X^3$. To solve the integral (\ref{external}) one may use
the Schwinger--Dyson equation technique \cite{BN81} written in terms of
eigenvalues of $\L $. The Kontsevich model was solved in the genus expansion
in the papers \cite{GN}, \cite{MakS} for genus zero (planar diagrams) and in
\cite{IZ92} for higher genera.

Recently, the Kontsevich--Penner model was introduced \cite{CM92a}.
The Lagrangian of this  model has the following form:
\be
{\cal Z}[\Lambda] = \int DX \exp \left( N \,\hbox{tr} \left\{
-\frac 12\Lambda X\Lambda X
+\alpha \bigl[ \log(1+X)-X\bigr]\right\}\right),\,\ \ \Lambda =
\hbox{ diag} (\Lambda_1, \dots, \Lambda_N).
\label{PK}
\ee
This model may
be readily reduced to (\ref{external}) with $V_0(X)= -X^2/2+\alpha \log X$.
It was solved in genus expansion in \cite{CM92a}, \cite{ACM}. It was shown in
\cite{CM92b}, \cite{KMMM} that it is in fact equivalent to the one--matrix
hermitean model (\ref{herm}) with the general potential
\be
P(X)=\sum_{n=0}^{\infty}T_n \tr X^n,
\ee
which times are defined by the kind of Miwa transform:
\be
T_n=\frac 1n \tr\L ^{-n} -\frac N2 \delta _{n2}\ \hbox{ for }\ n\ge 1
\ \ \hbox{ and }\ T_0=\tr \log \L ^{-1}.
\ee
Thus the Kontsevich--Penner model may be treated as an intermediate link
between 2D gravity described by the Kontsevich model and random surface
technique.

Indeed, as we shall demonstrate, this new model describes in a very natural
way the Kontsevich indices for the case of discretized moduli space If
we do not cast the moduli space boundary effects (which are relevant in
this model), then the coefficients of expansion are just the Kontsevich
indices, but since we deal with the closure of the moduli space, the answer
is tuned in a way to incorporate the Riemann surfaces which are boundary
components under the Deligne and Mumford reduction procedure \cite{Mum83}.
Then when taking the scaling limit (but keeping $N$ finite), we get just the
Kontsevich model. From the other hand, taking another rescaling we obtain
the Penner model describing virtual Euler characteristics of the moduli space
via the cell decomposition. Thus, this model provides a bridge between
Harer, Zagier and Penner theory \cite{HZ86}, \cite{Pen86} describing
virtual Euler characteristics on moduli space and the Kontsevich theory
giving intersection numbers of stable cohomology classes on the moduli space.

\newsection{The geometric approach to the Kontsevich model.}

In his original paper \cite{Kon91} Kontsevich proved that
\be
\sum_{d_1,\dots,d_s=0}^{\infty}<\!\tau_{d_1},\tau_{d_2},\dots,\tau_{d_s}\!>
\prod_{i=1}^{s}(2d_i-1)!!\l _i^{-(2d_i+1)}=
\sum_{\Gamma}{2^{-n_0}\over \#\hbox{\,Aut\,}(\Gamma )}
\prod_{\{ij\}}{2\over \l_i+\l_j},
\label{Aut}
\ee
where the objects standing in angular brackets on the left--hand side are
(rational) numbers describing intersection indices, and on the right--hand
side the sum runs over all oriented connected trivalent ``fatgraphs''
$\Gamma$ with $s$ labeled boundary components, regardless of the genus,
$n_0$ is the number of vertices of $\Gamma$, the product runs over all the
edges in the graph and $\#\hbox{\,Aut}$ is the volume of discrete symmetry
group of the graph $\Gamma$.

The amazing result by Kontsevich is that the quantity on the right hand
side of (\ref{Aut}) is equal to a free energy in the following matrix
model:
\be
\e^{F_N(\L)}={\int dX\,\exp\left(-\frac 12 \tr \L X^2+\frac i6 \tr X^3\right)
\over \int dX\,\exp\left(-\frac 12 \tr \L X^2 \right)},
\label{Konts}
\ee
where $X$ is an $N\times N$ hermitian matrix and $\L=\hbox{\,diag\,}(\l_1,
\dots ,\l_N)$. The distinct feature of the expression (\ref{Aut}) is that
in spite of the fact that each selected diagram has quantities
$(\l_i+\l_j)$ in the
denominator, when taking a sum over all diagrams of the same genus and
the same number of boundary components all these quantities are cancelled
with the ones from nominator.

Feynman rules for the Kontsevich matrix model are the following: as in the
usual matrix models, we deal with so-called ``fat graphs'' or ``ribbon
graphs'' with propagators having two sides, each carries corresponding index.
The Kontsevich model varies from the standard one--matrix hermitian model
since there appear additional variables $\l_i$ associated with index
loops in the diagram, the propagator being equal to $2/(\l_i+\l_j)$, where
$\l_i$ and $\l_j$ are variables of two cycles (perhaps the same cycle)
which the two sides of propagator belong to. Also there are trivalent
vertices presenting the cell decomposition of the moduli space.

It is instructive to consider the simplest example
of genus zero and three boundary components which we symbolically label
$\l_1$, $\l_2$ and $\l_3$. There are two kinds of diagrams giving the
contribution in this order (Fig.1).
The contribution to the free energy arising from this sum is
\bea
&{}& {1\over 6(\l_1+\l_2)(\l_1+\l_3)(\l_2+\l_3)}+\frac 13 \biggl\{
{1\over 4\l_1(\l_2+\l_1)(\l_3+\l_1)}\biggr. + \nonumber\\
&+&\biggl.(1\ra 2,2\ra 3, 3\ra 1)+
(1\ra 3,3\ra 2,2\ra 1)\biggr\}\nonumber\\
&=&{2\l_1\l_2\l_3+\l_2\l_3(\l_2+\l_3) +\l_1\l_3(\l_1+\l_3)
+\l_1\l_2(\l_1+\l_2)\over 12\l_1\l_2\l_3(\l_1+\l_2)(\l_1+\l_3)(\l_2+\l_3)}
\nonumber\\
&=&{1\over 12\l_1\l_2\l_3}.
\eea
This example demonstrates the cancellations of $(\l_i+\l_j)$--terms in
the denominator above mentioned.

%---------------------------------
% Figure 1 -- the g=0, s=3 contribution to Kontsevich's model
%
%           ______
%       #1 / ____ \            ___          ___
%         / /#2  \ \          / _ \  #1    / _ \
%        / /______\ \        / / \ \______/ / \ \
%   1/6 <  ________  >+ 1/2 < <#2 > ______ <#3 > > + perm.
%        \ \ #3   / /        \ \_/ /      \ \_/ /
%         \ \____/ /          \___/        \___/
%          \______/
%
%----------------------------------------------

\vspace{5cm}

\centerline{Figure 1.  the g=0, s=3 contribution to Kontsevich's model}

\vspace{6pt}

Now the sketch of Kontsevich's proof is in order. Let us  associate  with
each edge $e_i$ of a fat  graph  its  length  $l_i>0$.  We  consider  the
orbispace $\Mcomb$ of fat graphs with all possible lengths of edges and
arbitrary valencies of vertices. Two graphs are equivalent if an
isomorphism between them exists. Let us introduce an important object ---
the space of $(2,0)$--meromorphic differentials $\omega (z)dz^2$ on a
Riemann surface with $g$ handles and $n$ punctures, the only poles of
$\omega (z)$ are $n$ double poles placed in the points of punctures
with strictly positive quadratic residues $p_i^2>0$, $(i=1,\dots,n)$. It
is Strebel's theorem \cite{Streb} which claims that the natural mapping
from $\Mcomb$ to the moduli space $\MgnR$, where ${\bf R}_+^n$ is the space
of residues, $p_i>0$ being perimeters of cycles, is
homeomorphism. Thus, varying $l_j$ and taking the composition of all
graphs we span the whole space $\MgnR$.

Each cycle can be interpreted as a boundary components $I_i$ of the
Riemann surface since in the Strebel metric it can be presented as
half-infinite cylinder with the puncture point placed at infinity. The
boundary of it consists of a finite number of intervals (edges). We
consider a set of line bundles ${\cal L}_i$ which fiber at a point
$\Sigma\in\Mgn$ is the cotangent space to the puncture point
$x_i$ on the surface $\Sigma$. The
first Chern class of the line bundle ${\cal L}_i$ admits a representation in
terms of the lengths of the intervals $l_j$. The
perimeter of the boundary component is $p_i=\sum_{l_\alpha\in I_i}l_\alpha$
and
\be
c_1({\cal L}_i)=\sum_{{a,b\in I_i\atop a<b}}d\left({l_a\over p_i}\right)\wedge
d\left({l_b\over p_i}\right),
\label{2form}
\ee
where the cyclic ordering is asssumed. Following Kontsevich we introduce
2--form $\Omega$:
\be
\Omega = \sum_{i=1}^{n}\frac 12 p_i^2 c_1({\cal L}_i).
\label{omega}
\ee
The intersection indices are generated by the integrals over appropriate
power $d=3g-3+n$ of the form $\Omega$:
\bea
\frac{2^d}{d!} \int_{\Mgn} \Omega ^d &=& \frac{1}{d!}
\int_{\Mgn}\bigl(p_1^2c_1({\cal L}_1)+\dots +
p_n^2c_1({\cal L}_n)\bigr)^{d}=\nonumber\\
&=&\sum_{\sum d_i=d}\prod_{i=1}^{n}{p_i^{2d_i}\over d_i!}<\tau_{d_1}\dots
\tau_{d_n}>_{g}.
\label{index}
\eea

One important note is in order. It is a theorem by Kontsevich that these
integrations extend continuously to the closure of the moduli space $\Mc$
following the procedure by Deligne and Mumford \cite{Mum83}, and the proper
integration goes over $\Mc \times {\bf R}^n_+$. (It means that we deal with a
stable cohomological class of curves.)

Taking the Laplace transform over variables $p_i$ we get:
\be
\int_{0}^{\infty}dp_i\e ^{-p_i\l_i}p_i^{2d_i}=(2d_i)!\l _i^{-2d_i-1}
\ee
for the quantities standing on the right--hand side of (\ref{index}). On
the left--hand side we have
\be
\int_{0}^{\infty}\dots \int_{0}^{\infty}dp_1 \wedge\dots\wedge dp_n \e
^{-\sum p_i \l_i}\int _{\Mgn }\e^{\Omega},
\ee
and due to cancellations of all $p_i^2$ multipliers with $p_i$'s in
denominators of the form $\Omega$ we get:
\be
\e ^{\Omega} dp_1\wedge\dots\wedge dp_n = c \prod_{\alpha} \wedge
dl_{\alpha}.
\label{volume}
\ee
Surprisingly, the constant $c$ does depend only on Euler characteristic
of the graph $\Gamma$, $c=2^{-\kappa}$, $\kappa=2-2g=$ \# vertices
--- \# edges $+n$. Thus we have
\be
\sum_{\sum d_i=d}\prod_{i=1}^{n}{(2d_i-1)!!\over \l_i^{2d_i+1}}
<\tau_{d_1}\dots\tau_{d_n}>_{g} = \sum _{\Gamma}{2^{-\kappa}\over \#
\hbox{\,Aut\,}\Gamma}\int [dl]\exp \left( \sum_{\alpha}l_{\alpha}
(\l_{\alpha}^{(1)}+\l_{\alpha}^{(2)})\right).
\ee
Here $\l_{\alpha}^{(1)}$ and $\l_{\alpha}^{(2)}$ are variables of two
cycles divided by $\alpha$th edge.
Integration over all $dl_{\alpha}$ gives us eventually the relation
(\ref{Aut}).

\newsection{The Penner--Kontsevich model.}

Now let us turn to the case of the Penner--Kontsevich model (PK model).
It includes
in variance with the Kontsevich model all powers of $X^n$ in the potential
since it describes the partition of moduli space into cells of a simplicial
complex, the sum running over all simplices with different dimensions. (On
the language of the Kontsevich model the lower the dimension, the more and
more edges of the fat graph are reduced). Then the virtual Euler
characteristic is obtained by weighting the simplices by
\be
{(-1)^{d_F}\over |G_F|}
\label{euler}
\ee
where $|G_F|$ denotes the order of a stabilizer of the subgroup of the
mapping class group, that is, the order of the symmetry group of the
corresponding fat graph. It is the Penner model which gives the answer for
the sum over $F$ in (\ref{euler}) as a free energy for a matrix model
\cite{Pen86},\cite{DV90}:
\be
\sum_{F}N^{2-2g}t^{2-2g-n}{(-1)^{d_F}\over |G_F|}=\log \int dX
\e^{Nt\,\tr[\log(1-X)+X]},
\label{Penner}
\ee
where $n$ is a number of punctures on genus $g$ Riemann surface.
Expansion of the free energy of this model in $N$ and $t$ reveals logarithmic
corrections which is a feature of $c=1$ theories.

We find the Feynman rules for the Kontsevich--Penner
theory (\ref{PK}). First, as in the standard Penner model, we have
vertices of all orders in $X$. Due to rotational symmetry, the factor $1/n$
standing with each $X^n$ cancels, and only symmetrical factor (\ref{euler})
survives. Also there is a factor $(-\alpha)$ standing with each vertex. As in
Kontsevich model, there are variables $\l_i$ associated with each cycle.
But the form of propagator changes --- instead of $2/(\l_i+\l_j)$ we have
$1/(\l_i\l_j+\alpha)$.

Let us consider the same case ($g=0$, $n=3$) as for Kontsevich model. One
additional diagram resulting from vertex $X^4$ arises and gives the
contribution with opposite sign (Fig.2).

\vspace{5cm}

%---------------------------------
% Figure 2 -- the g=0, s=3 contribution to Penner-Kontsevich's model
%
%           ______
%       #1 / ____ \            ___          ___           ___ #1 ___
%         / /#2  \ \          / _ \  #1    / _ \         / _ \  / _ \
%        / /______\ \        / / \ \______/ / \ \       / / \ \/ / \ \
%   1/6 <  ________  >+ 1/2 < <#2 > ______ <#3 > > -1/2< <#2 >  <#3 > >+perm.
%        \ \ #3   / /        \ \_/ /      \ \_/ /       \ \_/ /\ \_/ /
%         \ \____/ /          \___/        \___/         \___/  \___/
%          \______/
%
%----------------------------------------------
\centerline{Fig.2. g=0, s=3 contribution to Penner--Kontsevich's model.}

\vspace{6pt}

This contribution is (symmetrized over $\l_1$, $\l_2$ and $\l_3$):
\bea
&-&\frac 13\left\{{\alpha\over 2(\l_1\l_2+\alpha)
(\l_1\l_3+\alpha)}-\hbox{\ perm.\ }\right\}
+ {\alpha ^2\over 6(\l_1\l_2+\alpha)(\l_1\l_3+\alpha)
(\l_2\l_3+\alpha)}\nonumber\\
&+&\frac 13 \left\{{\alpha ^2\over
2(\l_1^2+\alpha)(\l_1\l_2+\alpha)(\l_1\l_3+\alpha)}
+\hbox{\ perm.\ }\right\}
\label{111}
\eea
Again collecting all terms we get:
\be
{\alpha\over 6\prod_{i<j}(\l_i\l_j+\alpha)}\left\{-\sum_{i<j} \l_i\l_j
-2\alpha+\alpha\left({\l_2\l_3+\alpha\over \l_1^2+\alpha}+
{\l_1\l_2+\alpha\over \l_3^2+\alpha}+
{\l_1\l_3+\alpha\over \l_2^2+\alpha}\right)\right\},
\label{112}
\ee
and after a little algebra we obtain an answer:
\be
F_{0,3}=\alpha {-\l_1\l_2-\l_1\l_3-\l_2\l_3+\alpha\over 3(\l_1^2+\alpha)
(\l_2^2+\alpha) (\l_3^2+\alpha)}.
\label{f03}
\ee
We see that here, just as in standard Kontsevich model, the cancellation
of intertwining terms in the denominator occures that leads to factorization
of the answer over $1/(\l_i^2+\alpha)$--terms. It should reveal an
underlying geometric structure of the model under consideration, where
quantities like (\ref{Aut}) are expected to arise.

\newsection{Relation between KP model and discretized moduli space.}

Now we turn to the description of underlying differential--geometric
structure of the Kontsevich--Penner model. Let us consider the case of
a discretized moduli space. Its description can be done most properly in terms
of lengths of edges constituting boundary components of the
surface. We assume all these lengths to be integers (probably zeros)
scaled by some factor $\ep$. Thus, perimeters $p_i$ belong now to $\ep\cdot
{\bf Z}_+$. Then we can present the first Chern class of the line bundle
${\cal L}_i$ in the same form as above
\be
c_1({\cal L}_i)=\sum_{a,b\in I_i\atop a<b}d\left({n_a\over p_i}\right)\wedge
d\left({n_b\over p_i}\right),
\label{disc2form}
\ee
where $dn_a$ are symbols satisfying standard relations, $dn_a$ lie in the
cotangent space to the continuous moduli space taken at a point
$(n_1,\dots ,n_{6g-6+3n})$ which belongs also to the discretized moduli
space. The action of the
external derivative $d$ follows the same rules as in continuous case and the
integration is replaced by a discrete half-infinite sum. The 2--form $\Omega$
is defined by the same formula (\ref{omega}), thus the ``volume formula''
(\ref{volume}) is also preserved. Subtleties appear when one has to integrate
over discrete moduli space $\Mdisc$. First, there are points in $\Mdisc$
which do not lie on the boundary of the moduli space but in which some of the
lengths $n_a$ are equal zero. These points correspond to graphs containing
vertices of order greater than three. In the continuum limit we did not take
into account such graphs since they correspond to subdomains of lower
dimensions in the interior of the moduli space and because
the integration measure
is continuous we may neglect them. Now the situation changed and we should
cast these diagrams as well.

Second, now we should explicitly take into account curves which are reduced
by Deligne--Mumford procedure \cite{Mum83}. Fortunately, we can present an
explicit integration over the boundary $\partial \Mdisc$ of the discrete
moduli space because it expands into a sum over direct products of connected
components of lower genera and a number of additionally inserted punctures
corresponding to reduced handles of the surface. Doing all possible
reductions we span the whole closure of $\Mdisc$:
\be
\frac{2^d}{d!} \int_{\Mdisc} \Omega ^d = \frac{1}{d!}
\int_{\Mcdisc}\bigl(\sum_{i=1}^{n}p_i^2c_1({\cal L}_i)\bigr)^{d}-
\frac{1}{d!}\int_{\partial \Mcdisc}\bigl(\sum_{i=1}^{n}p_i^2c_1({\cal
L}_i)\bigr)^{d\p }.
\label{closure}
\ee
Here $d\p <d$ depends on the power of reduction.

A point $\SS\in\partial\Mdisc$ is a union of $s$ $(1\leq s\leq n+2g-2)$
connected components $\SS_{g_j,n_j,k_j}$, $(j=1,\dots, s)$. Each surface
$\SS_{g_j,n_j,k_j}$ has genus $g_j$, $\sum_{j=1}^{s}g_j\leq g$, $n_j$
original punctures $\left(\sum_{j=1}^{n}n_j=n\right)$ and
$k_j$ additional punctures
${\cal P}_l^{(j)}$ arising from the reduction procedure. The  linear  bundles
${\cal L}_i$ are associated with $n_j$ points of this surface but not with
the new ${\cal P}_l^{(j)}$. Explicitly the boundary $\partial \Mdisc$ can be
presented as a finite set of disconnected pieces, each of them is in its turn
the direct product of lower dimensional closed moduli spaces weighted with
$(-1)^{r_s}$, $r_s=\frac 12 \sum_{j}k_j$ being the reduction power:
\be
\otimes _{j=1}^{s} \Mdclose {g_j,n_j+k_j}(-1)^{r_s}.
\ee
Integration over $\Mdclose {g_j,n_j+k_j}(-1)^{r_s}$ expands into product of
integrals over connected closed components which are given by known
continuous Kontsevich's indices. So we conjecture the answer for intersection
indices on the discretized moduli space:
\bea
\frac{2^d}{d!} \int_{\Mdisc} \Omega ^d =
\sum_{\sum d_i=d}\,\prod_{i=1}^{n}{p_i^{2d_i}\over d_i!}<\tau_{d_1}\dots
\tau_{d_n}>_{g}\nonumber\\
+\sum_{reductions}(-1)^{r_s}{(d-r_s)!\over d!}
\prod_{j=1}^{s}\left( \sum_{\sum d_a=d_j}\,\prod_{a=1}^{n_j}
{p_a^{2d_a}\over d_a!}<\tau_{d_1}\dots
\tau_{d_{n_j}}\tau_0^{(1)}\dots \tau_0^{(k_j)}>_{g_j}\right),
\label{discindex}
\eea
where $<\dots >_{g_j}$ are corresponding Kontsevich's indices,
$d_j=3g_j-3+n_j+k_j$ and $k_j$ insertions of $\tau_0$ correspond to
additional punctures.

Next step is to do the Laplace transform over variables $p_i$. Now we should
sum over $p_i=\epsilon, 2\epsilon, 3\epsilon, \dots$ simultaneously taking
into account that $\sum_{i=1}^{n}p_i\in 2\epsilon {\bf Z}_+$ (every edge
$n_a$ is counted twice). A procedure is the following. We do the transform
over variables $\l_j$ taking a sum over all $p_j\in \epsilon {\bf Z}_+$
weighted with $(i)^j$ and release the real part of the obtained expression
consequently substituting $\e^{\epsilon \l_j}\to i\e ^{\epsilon\l_j}$. On the
right hand side we have (explicitly reconstructing the
$\epsilon$--dependence):
\be
\hbox{ Re}\,\prod_{j=1}^{n}\left(\sum_{p_j=1}^{\infty}\e^{-\epsilon \l_j p_j}
(i)^{p_j}\epsilon ^{2d_j}p_j^{2d_j}\right)=\prod_{j=1}^{n}\left(\dd {\l_j}
\right)^{2d_j}\,\hbox{ Re}{(-1)^n \over \prod_{j=1}^{n}(1+i\e^{\epsilon
\l_j})}.
\label{discLapl}
\ee
Taking as an example the case of ${\cal M}_{0,3}$ we immediately get the
expression (\ref{f03}).

On the right hand side we get:
\bea
&{}&\sum_{\{p_i\}\in\Mdisc} \e^{-\epsilon \l_j p_j} \epsilon ^n \e ^{\Omega}
dp_1\wedge \dots \wedge dp_n \nonumber\\
&=&\int _{\Mdisc}\e^{-\epsilon \l_j p_j} 2^{2g-2}dn_1\wedge \dots \wedge
dn_{6g-6+3n} \epsilon ^{6g-6+3n}.
\eea
The last term possesses an explicit representation as a sum over all possible
graphs with fixed genus $g$ and number of cycles $n$ and {\sl arbitrary}
valencies of vertices:
\bea
&{}&\sum_{\Gamma}{1\over \#\hbox{\,Aut\,}(\Gamma )}2^{\#edges-\#vert.+n}
\epsilon ^{6g-6+3n}\prod_{s}\sum_{n_s=1}^{\infty}\e^{-\epsilon n_s(\l_s^{(1)}
+\l_s^{(2)})}\nonumber\\
&{}&=2^n\sum_{\Gamma}{1\over \#\hbox{\,Aut\,}(\Gamma )}2^{-\#vert.}
\epsilon ^{6g-6+3n}\prod_{\{ij\}}{2\over
\e^{\epsilon (\l_i +\l_j)}-1}\nonumber\\
&{}&\equiv w_g(\l_1,\dots,\l_n).
\label{discAut}
\eea

This last expression  is in fact the free energy term for the
Kontsevich--Penner matrix model in fixed $g$ and $n$ of the form
\be
\e^{F_N(\Lambda)} = {\int DX \exp N\alpha\,\tr\,
\left\{-\frac 14\Lambda X\Lambda X
-\frac 12 \bigl[ \log(1-X)+X\bigr]\right\}\over
\int DX \exp N\alpha\,\tr\,
\left\{-\frac 14\Lambda X\Lambda X
+\frac 14 X^2 \right\}},
\label{PKdisc}
\ee
where $\alpha=1/\epsilon ^3$ and $\Lambda = \hbox{ diag}\,
(\e ^{\epsilon \l_1}, \e ^{\epsilon \l_2}, \dots, \e ^{\epsilon \l_N})$.

$F_N(\Lambda)$ has an expansion which looks just like (\ref{taufunct}):
\be
F_N(\Lambda) = \sum_{{g=0\atop n=1}}^{\infty}(N\alpha )^{2-2g}\alpha ^{-n}
N^{-n} \tr w_g(\l_1,\dots,\l_n).
\ee

In the continuum limit $\epsilon\to 0$ we immediately get the expression
(\ref{Konts}) i.e. the Kontsevich matrix model.

Thus we have demontrated the relation between geometric characteristics of
the discretized moduli space and related matrix models. We show how the
Kontevich indices can be naturally embedded into standard one--matrix model
via the Penner--Kontsevich model.

In conclusion we should note some problems and perspectives of the proposed
model. First, we did not yet prove rigorously the relation (\ref{discindex}),
and this proof is necessary for the completeness of the theory. Also it is
interesting to make the sum over reductions more explicitly (in the
combinatorial sense). Second interesting problem to solve is to find the
description of just the Penner model in the case of the discretized moduli
space. The formula (\ref{Penner}) should be modified since in this case the
volume of the stabilizer (symmetry group) taken at the boundary (infinite)
point might be no infinite but rather proportional to some positive power of
$\epsilon$, the discretization parameter. Also
it is interesting to develop this approach
to the case of GKM, where most subtle geometric invariants are considered.

I am grateful to G.Falqui, V.Fock, A.Gerasimov, Yu.Makeenko and A.Mironov
for numerous valuable discussions.

\end{document}